\documentclass[11pt,twoside]{article}
\usepackage{asp2010}

\resetcounters

\begin{document}

\title{On the seismic modelling of rotating B-type pulsators in the
traditional approximation}
\author{Conny Aerts$^1$, Marc-Antoine Dupret$^2$
\affil{$^1$Instituut voor Sterrenkunde, K.U.Leuven, Belgium; \\Dep.\ 
  Astrophysics, IMAPP, Radboud
  University Nijmegen, the Netherlands}
\affil{$^2$Institut d'Astrophysique et de G\'eophysique, Universit\'e de
  Li\`ege, Belgium}}

\begin{abstract} {The CoRoT and {\it Kepler\/} 
data revolutionised our view on stellar
    pulsation. For massive stars, the space data revealed the simultaneous
    presence of low-amplitude low-order modes and dominant high-order gravity
    modes in several B-type pulsators.  The interpretation of such a rich set of
    detected oscillations requires new tools.  We present computations of
    oscillations for B-type pulsators taking into account the effects of the
    Coriolis force in the so-called traditional approximation. We discuss the
    limitations of classical frequency matching to tune these stars seismically
    and show that the predictive power is limited in the case of high-order
    gravity mode pulsators, except if numerous modes of consecutive radial order
    can be identified.  }
\end{abstract}

\section{Current status and goal}

The status of observational asteroseismology across the HR diagram has changed
dramatically since the data of the CoRoT and {\it Kepler\/} space missions
became available (for a review of this status prior to CoRoT and Kepler, see
Chapter\,2 of Aerts et al.\ 2010). Regarding massive stars with slow rotation,
we have a fairly good understanding of B-star oscillations, particularly for the
$\beta\,$Cep stars whose oscillation spectra are dominated by low-order pressure
(p) and gravity (g) modes with periods of a few hours. Internal structure
parameters, such as core convective overshooting and core versus envelope
rotation, were already tuned from hugh multisite campaigns, for a few very
bright class members, by performing frequency matching of their zonal modes
after empirical mode identification from multi-colour photometry and/or
high-resolution spectroscopy.  Seismic modelling of Slowly Pulsating B Stars
(SPBs hereafter) was at a much lower level prior to the era of the space
photometry. This was a consequence of them being high-order g mode pulsators
with periodicities of days, which implies a challenge to detect a sufficient
number of oscillations in ground-based data.  As already hinted at by the +20
frequencies deduced from an uninterrupted 6-weeks time series of space
photometry of the slowly rotating B5 star HD163830 assembled with the MOST
satellite (Aerts et al.\ 2006), immense progress was to be expected for similar
stars from more precise data with a much longer time base to be assembled with
CoRoT and {\it Kepler\/}.

The CoRoT data of the B3V star HD\,50230 brought a new view on the seismic
modelling of slowly rotating SPBs. Degroote et al.\ (2010) not only found period
spacings from the 5-month CoRoT light curve of that star, but even periodic
deviations from a constant period spacing with decreasing amplitude as the mode
period increases as expected from theory for high-order g modes (Miglio et al.\
2008).  In order to interpret this detection, it was necessary to include a
chemically inhomogeneous near-core region in stellar models of {\it
  non-rotating\/} stars.  A natural cause for this necessary extra diffusive
mixing could be of rotational nature, but other origins cannot be excluded.

HD\,50230 is situated in the common part of the $\beta\,$Cep and SPB instability
strips. Pressure modes are indeed also detected in the CoRoT light curve, at
lower amplitude than the g modes (Degroote et al.\ 2010). This is also the
situation encountered for the pulsators of spectral type B observed by
the {\it Kepler\/} satellite (Balona et al.\ 2011), but period spacings have not
yet been reported for those.

Empirical mode identification of the detected g modes in the B pulsators found
from the white-light space photometry is hard, because these modes have
amplitudes of only a few mmag or less, implying that methods based on
ground-based multicolour photometry or high-precision spectroscopy are not
feasible. We are thus left with frequency or period matching or, preferrably,
with period spacings as adopted by Degroote et al.\ (2010) as a powerful
diagnostic for seismic modelling.

In this paper, we show that matching of the detected frequencies of g modes in
hybrid pulsators is a good way to tune the star, provided that modes of
consecutive radial order can be found and identified and that the Coriolis force
is taken into account.  Indeed, even though SPBs are generally very slow
rotators, their g-mode periods are of the same order than their rotation period,
which implies that one cannot ignore the Coriolis force in the computation of
their frequency spectra. Luckily, the slow rotation does allow to
neglect the effects of the centrifugal force in most cases, i.e., to consider a
spherically symmetric equilibrium configuration.

From a theoretical point of view, hybrid pulsators are stars which have both
unstable high-order g modes and low-order p and g modes, separated by a stable
region of intermediate-order g modes. In practice, however, the predicted gap in
the frequency spectra can easily disappear, as seems to be the case in many of
the hybrids observed with CoRoT and {\it Kepler\/}, because the rotational
splitting of modes with $\ell>2$ may cause shifts towards this ``empty'' region,
or the occurrence of negative g-mode frequencies in an inertial frame may be
wrongly interpreted from data analysis methods which
assume positive frequencies.
\begin{figure}[t!]
\begin{center}
\rotatebox{270}{\resizebox{8.cm}{!}{\includegraphics{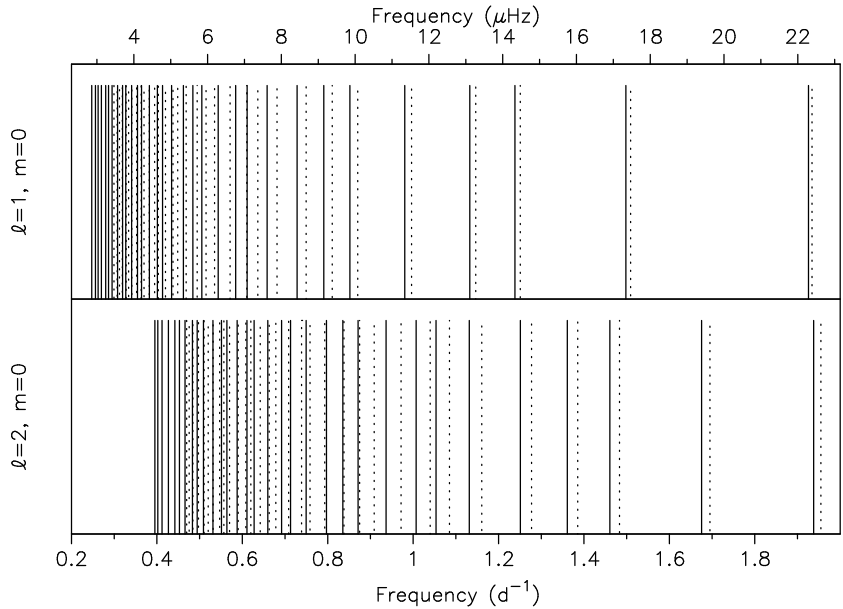}}}
\end{center}
\caption{Oscillation frequencies of zonal $\ell=1$, g$_{35}$ to g$_4$ modes
  (upper panel) and $\ell=2$, g$_{38}$ to g$_7$ modes (lower panel), computed in
  the traditional approximation for a stellar model halfway the
  main sequence phase having $M=7.5\,$M$_\odot$, $Z=0.015$, $X=0.70$,
  $\alpha_{\rm ov}=0.2$, for equatorial surface velocities of 10\,km\,s$^{-1}$
  (full lines) and 50\,km\,s$^{-1}$ (dotted lines).  }
\label{fig1}
\end{figure}
\begin{figure}[t!]
\begin{center}
\rotatebox{270}{\resizebox{8.cm}{!}{\includegraphics{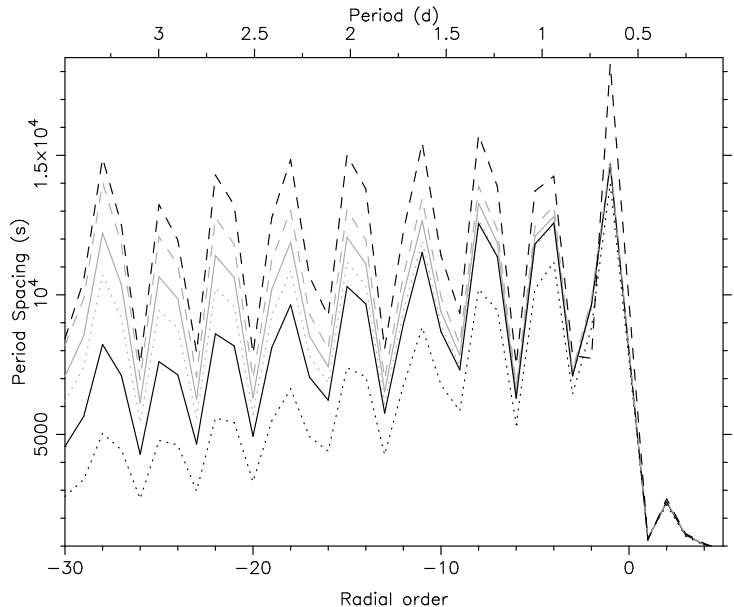}}}
\end{center}
\caption{Period spacings of $\ell=1$ modes for the model discussed in
  Fig.\,\protect\ref{fig1} and for equatorial surface velocities of
  10\,km\,s$^{-1}$ (grey) and 50\,km\,s$^{-1}$ (black). Dotted lines: $m=-1$,
  full lines: $m=0$, dashed lines: $m=+1$. }
\label{fig2}
\end{figure}
\begin{figure}[t!]
\begin{center}
\rotatebox{270}{\resizebox{8.cm}{!}{\includegraphics{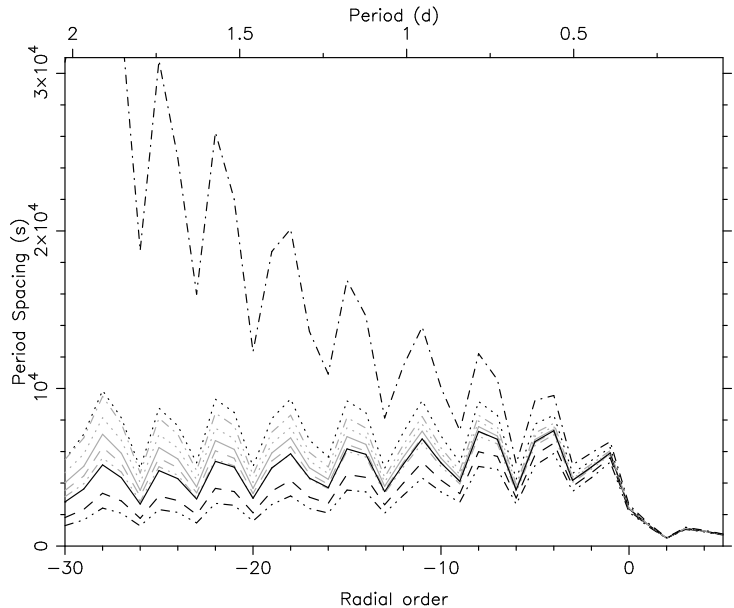}}}
\end{center}
\caption{Period spacings of $\ell=2$ modes for the model discussed in
  Fig.\,\protect\ref{fig1} and for equatorial surface velocities of
  10\,km\,s$^{-1}$ (grey) and 50\,km\,s$^{-1}$ (black). Dashed-dot-dot-dot
  lines: $m=-2$, dashed lines: $m=-1$, full lines: $m=0$, dotted lines: $m=+1$,
  dashed-dotted lines: $m=+2$. }
\label{fig3}
\end{figure}
\begin{figure}[t!]
\begin{center}
\rotatebox{270}{\resizebox{7.5cm}{!}{\includegraphics{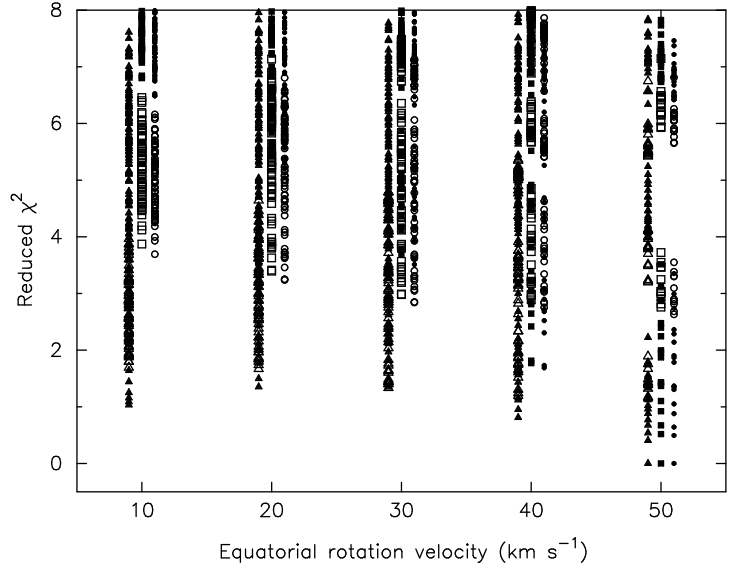}}}
\end{center}
\caption{Three versions of the reduced $\chi^2$ according to
  Eqs\,(\protect\ref{chi}), where triangles, squares, and circles correspond
  with $\chi_1^2$, $\chi_2^2$, and $\chi_3^2$, respectively. The symbols for
  $\chi_1^2$ and $\chi_3^2$ have been shifted slightly in equatorial rotation
  velocity for better visibility. Open/closed 
symbols denote models whose dipole/quadrupole modes
  match the ten observed ones.}
\label{fig4}
\end{figure}

\section{Gravity mode frequencies in the traditional approximation}

Given that we have by far the best observational constraints for the CoRoT B3V
target and hybrid pulsator HD\,50230 (Degroote 2010), we consider models
that are representative for this star.  However, our results remain
qualitatively valid for lower mass stars in the SPB instability strip as well.

The surface rotation velocity of SPBs is low, i.e., typically less than 20\% of
their critical velocity and for HD\,50230 only a few percent.  It is thus
justified to consider spherically symmetric equilibrium models.  We considered
non-rotating models to compute the equilibrium quantities, but we do take into
account the effects of the rotation in the computation of the perturbed
quantities due to the oscillations. The evolutionary models were computed with
the Code Li\'egois d'\'Evolution Stellaire (CL\'ES, Scuflaire et al.\ 2008) with
the same input physics as used and described by Briquet et al.\ (2011) and thus
not repeated here. Each of the models is characterised by a value for the mass
$M$, the initial hydrogen content $X$, the initial metallicity $Z$, the core
overshooting parameter $\alpha_{\rm ov}$ (a dimensionless quantity expressed as
a fraction of the local pressure scale height) and the age (or, equivalently,
the central hydrogen content $X_c$).

It is known since long that the effects of rotation on the g modes of SPBs are
well described by adopting the so-called traditional approximation. A thorough
recent discussion can be found in Townsend (2003), and references therein. In
the traditional approximation, one takes into account the effect of the Coriolis
force in the pulsation equations to compute the displacement field and the
frequency of each mode, under the assumptions that the horizontal component of
the rotation vector can be ignored and that the Brunt-V\"ais\"al\"a frequency is
much higher than the mode frequency.  This is fully justified for the g modes in
SPBs, whose horizontal component of the displacement field of the oscillations
is typically a factor 50 to 100 larger than the radial component and whose
oscillation frequencies are low. We are thus in the optimal regime to apply the
traditional approximation to the g modes. This approximation is less relevant
for p modes since they have a dominant radial displacement vector such that the
neglect of the horizontal component of the rotation vector with respect to the
one of the displacement vector is harder to justify, although it does not do any
``harm'' in the sense that the resulting frequencies and eigenfunctions will
have similar values compared with the case where the Coriolis force is ignored
altogether.  The traditional approximation leaves the radial modes unaffected
(see, e.g., Chapter 3 in Aerts et al.\ 2010).

One of us (MAD) adapted the non-adiabatic pulsation code MAD (Dupret 2001,
Dupret et al.\ 2002) to include rotational effects in the traditional
approximation, according to the formalism of Townsend (2003). In this
semi-analytical formalism, each of the mode eigenfunctions is written as a
finite sum of spherical harmonics. The angular mode quantum numbers are real
numbers in this case, but they converge to the usual integers $(\ell,m)$ in the
case of no rotation and can be identified as such, thus we keep using this
classical notation here.  The sign convention we adopted is $m<0$ for prograde
modes and $m>0$ for retrograde modes.  We considered low equatorial rotation
velocities of 10\,km\,s$^{-1}$ and 50\,km\,s$^{-1}$, which are typical for SPBs
(Aerts et al.\ 1999). For several main-sequence models, the g modes of radial
order 1 to 50 were computed, as well as the p modes of radial order from 1 to
10, for dipole as well as quadrupole modes, including all possible values for
the azimutal order $m=-\ell,\ldots,\ell$.

Parts of the zonal $\ell=1$ and $\ell=2$ g-mode frequency spectra for a typical
stellar model with parameters $M=7.5$\,M$_\odot$, $X=0.70$, $Z=0.015$,
$\alpha_{\rm ov}=0.2$ and an age of 28.1 million years (corresponding with
$X_C=0.342$) are shown in Fig.\,\ref{fig1}. It can be seen that the shift in
frequency, when increasing the equatorial rotation velocity from
10\,km\,s$^{-1}$ to 50\,km\,s$^{-1}$, is large, reaching 0.052\,d$^{-1}$ for the
dipole g$_{35}$ mode and 0.064\,d$^{-1}$ for the quadrupole g$_{38}$ mode. These
shifts are 0.009\,d$^{-1}$ and 0.018\,d$^{-1}$ for the $\ell=1$, g$_4$ and
$\ell=2$, g$_7$ mode, respectively. The frequency changes are below
0.001\,d$^{-1}$ in the p-mode regime and thus essentially less than the
theoretical uncertainty due to various different versions of oscillation
codes (Moya et al.\ 2008).

A typical CoRoT data set leads to a Rayleigh limit for the frequency resolution
of about 0.006\,d$^{-1}$ while the {\it Kepler\/} data sets will reach better
than 0.0004\,d$^{-1}$ for the nominal life time of the mission. It is thus
suggested by Fig.\,\ref{fig1} that one should not perform seismic modelling by
``blind'' frequency matching of g-mode frequencies alone while ignoring
rotational effects, i.e., the inclusion of the Coriolis force as well as
(partial) identification of the mode quantum numbers $(\ell,m)$ are necessary
to achieve reliable results, along with the detection of frequency {\it
  patterns}.

\section{Seismic diagnostic tools}

The effect of rotation on the instability of g modes in B stars in the
traditional approximation was investigated by Townsend (2005a,b) to which we
refer for details and other references.  We only considered one value for the
mass representing HD\,50230 here.  In agreement with Townsend's results, we find
that the mode excitation is hardly affected compared to the case where the
rotation is ignored in the pulsation computations for the low rotational
velocities we consider in this paper. For the model whose spectrum
is shown in Fig.\,\ref{fig1}, we find instability for the radial orders -21 to
-12 for the zonal quadrupole modes, and only for the zonal dipole mode of order
-1.

Figs\,\ref{fig2} and \ref{fig3} show the period spacings of the g and p modes of
$\ell=1$ and $\ell=2$ for radial orders from -30 to 5 for the stellar model
whose frequency spectra were partly shown in Fig.\,\ref{fig1}. The relevant mode
periods have also been indicated to guide comparison with observations.  First
of all, we recover the well-known result that period spacings are not as
suitable a diagnostic for the p modes than for g modes.  Secondly, the shape of
the period spacings for the zonal modes in the case of slow equatorial rotation
velocity is similar to the one obtained by Miglio et al.\ (2008) who made a very
extensive theoretical study of such quantities in the absence of rotation, for
both SPB and $\gamma\,$Dor models. Thirdly, an asymmetry occurs in the way the
period spacings are changed by the rotation in the sense that those of
retrograde modes are much more affected than those of prograde modes.

Even though it is not our preferred way of working, frequency matching is a
common practise in seismic modelling.  In order to test how much the Coriolis
force affects such matching procedure when one compares detected frequencies
with those predicted for a wrong rotation velocity, we performed several
experiments.  We generated a frequency spectrum for an equatorial rotation
velocity of 50\,km\,s$^{-1}$ and we selected ten excited quadrupole retrograde
sectoral mode frequencies of consecutive radial order from -21 to -13, having
frequencies from 0.37 to 0.72\,d$^{-1}$ in an inertial frame, which we consider
as the ``observed'' frequency set $f_{\rm obs}$, for a model with fixed $(M, X,
Z, \alpha_{\rm ov})$ on the main sequence. Subsequently, we scanned the
pulsation spectra of 35 models along this evolutionary track, computed for
$\ell=1$ and 2 in the traditional approximation, for a range of equatorial
rotation velocities from 10 to 50\,km\,s$^{-1}$, in steps of 10\,km\,s$^{-1}$,
assuming that we know it concerns ten modes of the same $(\ell,m)$ of
consecutive radial order, without knowing the $(\ell,m)$.  These circumstances
will in general be too optimistic compared to reality, but the case of HD\,50230
in Degroote et al.\ (2010) shows that it is possible to detect such a number of
modes constituting a series of radial order, with a period spacing and a
deviation thereof, which comes close to our experimental set-up except that we
considered a star with a larger rotational velocity.  The experiment allows us
to check the effect of the Coriolis force alone, without being affected by the
unknown evolutionary track.

We requested a match between the highest ``observed'' frequency and the model
frequencies $f_{\rm mod}$ better than 0.02\,d$^{-1}$, which is an estimate of
the theoretical frequency uncertainty (Moya et al.\ 2008). We also computed the
average period spacing $\langle\Delta\,P\rangle$ of these ten modes, as well as
the average frequency spacing $\langle\Delta\,f\rangle$.  For all the models and
$(\ell,m)$ sets that fulfill these requirements, we computed three different
versions of a reduced $\chi^2$:
\begin{equation}
\renewcommand{\arraystretch}{3.5}
\begin{array}{l}
\chi_1^2\equiv\!\sqrt{
\displaystyle{
\frac{1}{9}\left[
\sum_{i=1}^{10}\frac{(f_{\rm obs}-f_{\rm mod})^2}{\sigma_{\rm f}^2}
\right]
}},\\
\chi_2^2\equiv\!\sqrt{
\displaystyle{
\frac{1}{10}\left[
\sum_{i=1}^{10}\frac{(f_{\rm obs}-f_{\rm mod})^2}{\sigma_{\rm f}^2}+ 
\frac{(\langle\Delta\,P\rangle_{\rm obs}-\langle\Delta\,P\rangle_{\rm mod})^2}
{(500\,{\rm sec})^2}
\right]
}},\\
\chi_3^2\equiv\!\sqrt{
\displaystyle{
\frac{1}{11}\left[
\sum_{i=1}^{10}\frac{(f_{\rm obs}-f_{\rm mod})^2}{\sigma_{\rm f}^2}+ 
\frac{(\langle\Delta P\rangle_{\rm obs}-\langle\Delta P\rangle_{\rm mod})^2}
{(500\,{\rm sec})^2}+
\frac{(\langle\Delta f\rangle_{\rm obs}-\langle\Delta f\rangle_{\rm mod})^2}
{4\sigma_{\rm f}^2}
\right]
}},
\end{array}
\label{chi}
\end{equation}
where we took $\sigma_{\rm f}=0.02$\,d$^{-1}$. By comparing the values of these
three merit functions, we can decide whether or not the addition of period and
frequency spacings helps in the selection of the best models when we are dealing
with g modes.

The results of this first experiment are shown in Fig.\,\ref{fig4}. First of
all, we recover the input model as the best one, as should be the case. Further,
we see that most of the best fitting models 
have quadrupole modes fitting the observed
ones, which implies one can in principle distinguish between dipole and
quadrupole modes from the values of the frequencies when we observe consecutive
radial orders. Another result is that we derive ``appropriate'' models (i.e.,
with $\chi^2<2$) for ``wrong'' equatorial rotation velocities.  Thus, it is not
possible to deduce the correct equatorial velocity when the star has sectoral
g modes without additional information.  The inclusion of a value for the average
period and frequency spacing in the $\chi^2$ usually leads to a slightly better
fit compared to the $\chi^2$ in which these quantities are ignored, but not
always since the average of ten detected spacings is not necessarily well
represented by the average of the model frequences.  Finally, for all models
with $\chi^2$ below 2, which is commonly adopted as the cut-off value to accept
a model or not, the ranges of the effective temperature, logarithm of the
gravity, and central hydrogen fraction are $[18000,20900]$\,K, $[3.59,4.02]$,
and $[0.076,0.414]$, respectively, compared with the input values of 20500\,K,
3.94, and 0.342. The predictive power to tune those is thus limited. Also, the
entire range of $[10,50]$km\,s$^{-1}$ occurs for the equatorial rotation
velocity of these models with $\chi^2<2$.

In a second experiment, we checked if these results remain similar
for a B-type pulsator with modes of lower radial order.
For the
same stellar model as in the first experiment, we now considered a series of ten
consecutive quadrupole zonal modes with radial orders ranging from -6 to 3,
having inertial frame frequencies from 2.12 and 13.85\,d$^{-1}$, for an
equatorial rotation velocity of 30\,km\,s$^{-1}$. In this case, only the input
model among the 35 models along the evolutionary track considered in the
frequency matching meets the requirements of having $\chi^2<2$ while the other
34 models have $\chi^2>30$, and the match for the surviving model is only
acceptable for the correct identification of $(\ell,m)=(2,0)$. In all cases, the
addition of the average period and frequency spacing in the $\chi^2$ leads to a
lower value, i.e., $\chi^2_3<\chi^2_2<\chi^2_1$. None of the  
$\chi^2$ functions is able to discriminate among the considered values for
the rotation velocity based on the criterion $\chi^2<2$.

A third and final experiment is the same as the second one, except that we took
the $(\ell,m)=(2,-1)$ modes of orders from -6 to 3, and an equatorial rotation
velocity of 40\,km\,s$^{-1}$. Also in this case the discriminating power among
the models is good and only two models fulfill our criteria: the input
model with the input modes and the correct rotation velocity, and the input
model with the $(2,-2)$ modes and half of the input rotation. All other models
again have very high $\chi^2$ values.

\section{Discussion}

The results obtained here imply good potential to model pulsating B stars
seismically, by adopting the traditional approximation, provided that we are
dealing with slow rotators with excited moderate to low-order modes. The case is
more worrisome for high-order mode pulsators, except if we can derive precise
values for period spacings based on numerous modes of consecutive radial order,
as in Degroote et al.\ (2010).  The combination of fitting the frequencies along
with a value for a period and/or frequency spacing offers a good way to select
appropriate models but the uniqueness of the selection is hard to prove.  The
prospects are good for pulsators with low-order p and g modes of various radial
orders if modes of the same $(l,m)$ can be detected and identified.

In a future study, this work will be generalised to an extensive grid of models
covering the entire instability strip of B pulsators, in order to study if the
matching of frequencies, along with period and frequency separations, is
sufficient to differentiate among various models with different $(M, X, Z,
\alpha_{\rm ov}, X_C)$ and equatorial rotation velocity instead of only the
latter two parameters as we considered here. Applications to observed stars will
then also be made as an improvement to presently available studies, where
seismic modelling was done while ignoring the rotational effects, if at all.

\acknowledgements The research leading to these results has received funding
from the European Research Council under the European Community's Seventh
Framework Programme (FP7/2007--2013)/ERC grant agreement n$^\circ$227224
(PROSPERITY). C.A.\ acknowledges the Fund for Scientific Research -- Flanders
for financial support to undertake a 6-month sabbatical leave.


\end{document}